\documentclass[12pt]{iopart}
\usepackage{iopams,graphicx,subfigure,color,mathptmx}  
\newcommand{\re}{\mathop{\mathrm{Re}} \nolimits}
\newcommand{\im}{\mathop{\mathrm{Im}} \nolimits}
\newcommand{\openone}{\leavevmode\hbox{\small1\normalsize\kern-.33em1}}
\newcommand{\sugg}[1]{{#1}}

\begin{document}

\title{$SU(1,1)$ covariant $s$-parametrized maps}
\author{Andrei~B~Klimov$^{1,*}$, Ulrich~Seyfarth$^{2}$, Hubert~de Guise$^{3}$ and Luis~L~S\'{a}nchez-Soto$^{2,4}$}

\begin{abstract}
We propose a practical recipe to compute the ${s}$-parametrized maps for
systems with $SU(1,1)$ symmetry using a connection between the ${Q}$ and ${P}
$ symbols through the action of an operator invariant under the group. The
particular case of the self-dual (Wigner) phase-space functions, defined on
the upper sheet of the two-sheet hyperboloid (or, equivalently, inside the
Poincar\'{e} disc) are analyzed.
\end{abstract}

\address{$^{1}$ Departamento de F\'{\i}sica, Universidad de Guadalajara,
44420~Guadalajara, Jalisco, Mexico}

\address{$^{2}$ Max-Planck-Institut f\"{u}r die Physik des Lichts, 
Staudtstra\ss e 2, 91058~Erlangen, Germany}

\address{$^{3}$ Department of Physics, Lakehead University, Thunder Bay,
Ontario P7B 5E1, Canada}

\address{$^{4}$ Departamento de \'{O}ptica, Facultad de F\'{\i}sica,
Universidad Complutense, 28040~Madrid, Spain}

\ead{klimov@cencar.udg.mx} \vspace{10pt} 
\begin{indented}
\item[\today]
\end{indented}

\vspace{2pc} \noindent\textit{Keywords}: $SU(1,1)$, Wigner function, Phase-space methods

\submitto{\JPA}

\eqnobysec

\section{Introduction}

Phase-space approaches often unveil hidden facets of quantum systems and
shed light on their underlying kinematical and dynamical properties~\cite{Hillery:1984aa,Lee:1995aa,Schroek:1996aa,Ozorio:1998aa,Schleich:2001aa,QMPS:2005aa,Polkovnikov:2010aa,Weinbub:2018aa}. This type of analysis is now common in many areas, especially for systems with Heisenberg-Weyl~\cite{Glauber:1963aa,Sudarshan:1963aa,Agarwal:1968aa,Cahill:1969aa,Agarwal:1970aa,Gadella:1995aa} or ${SU}(2)$ symmetries \cite{Agarwal:1981aa,Varilly:1989aa}, and has been extended to other dynamical groups \sugg{such as} ${SU}(N)$~\cite{Klimov:2010aa,Tilma:2016aa} or ${E}(2)$~\cite{Gadella:1991aa,Nieto:1998cr,Plebanski:2000fk,Kastrup:2006cr,Rigas:2011by,Kastrup:2016aa}.

Following the pioneering work of Moyal~\cite{Moyal:1949aa}, \sugg{Groenewold~\cite{Groenewold:1946aa}} and Stratonovich~\cite{Stratonovich:1956aa}, the states of a quantum system in the Hilbert space $\mathcal{H}$ that carries an irreducible representation (irrep) $\Lambda $ of a dynamical group $G$ can be mapped into functions of a classical phase-space $\mathcal{M}$, wherein $G$ acts transitively. The structure of the manifold $\mathcal{M}$ is closely related to a set of coherent states $\{|\zeta \rangle \}$ labelled with phase-space coordinates $\zeta \in \mathcal{M}$~\cite{Onofri:1975aa}.

When coherent states can be constructed as translates of a fixed cyclic vector \cite{Perelomov:1986ly,Zhang:1990aa,Gazeau:2009aa} two mutually dual maps are naturally defined: they put in correspondence each operator $\hat{A}$ acting in the Hilbert space of the quantum system, with the so-called ${Q}$ and ${P}$ symbols, respectively, defined as~\cite{Husimi:1940aa,Kano:1965aa,Berezin:1975mw} 
\begin{equation}
{Q}_{A}(\zeta )=\langle \zeta |\hat{A}|\zeta \rangle \,,\qquad \qquad \hat{A}
=\int \rmd \mu (\zeta )\;P_{A}(\zeta )\;|\zeta \rangle \langle \zeta |\,,
\label{QP}
\end{equation}
where $d\mu (\zeta )$ is the normalized invariant measure on $\mathcal{M}$.
These symbols allow the computation of average values as a convolution 
\begin{equation}
\Tr(\hat{A}\hat{\varrho})=\int \rmd\mu (\zeta )\,P_{A}(\zeta )Q_{\varrho
}(\zeta )\,,  \label{av QP}
\end{equation}
with $\hat{\varrho}$ the density operator for the system.

\sugg{In theory, $Q$- and $P$-maps are both exact and contain complete information about the system. In practice, however, they are not always suitable for the analysis of quantum correlations. In particular, the $P$-symbols may become singular, whereas the $Q$-symbols are too smooth and do not exhibit the full quantum interference pattern. Moreover, in the semiclassical limit, the description of the dynamics in terms of the $P$- and $Q$-functions is not always appropriate: the corrections are of first order in the expansion  parameter (whose form is dictated by the symmetry of the system), which may lead to a considerable reduction of the timescale over which the semiclassical approximation is valid.}

\sugg{The Wigner map, $\hat{A} \leftrightarrow W_{A}(\zeta )$, is free of these difficulties. It satisfies}
\begin{equation}
\Tr(\hat{A} \hat{\varrho}) = \int \rmd\mu (\zeta )\;
W_{A}(\zeta )\,W_{\varrho}(\zeta ) \,.  
\label{sdW}
\end{equation}
The Wigner symbol of the density matrix (the so-called Wigner function) is not singular (for physical states), and has been shown to be very useful for analysis of the quantum states both in the deep quantum and semiclassical limits~\cite{Klimov:2017aa,Valtierra:2017aa}.

More generally one can introduce a  \sugg{parametrized family} of trace-like maps generated by kernels $\hat{w}^{(s)}\left( \zeta \right) $ 
\begin{equation}
{W}_{A}^{(s)}(\zeta )=\Tr [ \hat{A} \, \hat{w}^{(s)}( \zeta ) ]\,,
\label{Ws}
\end{equation}
where the parameter $s$ has an explicit interpretation in terms of ordering for the Heisenberg-Weyl algebra, with $\pm 1$, $0$ associated with ${P}$-, ${Q}$- and Wigner maps respectively~\cite{Cahill:1969aa}. \sugg{The same kind of mapping exists for higher symmetries, albeit the parameter $s$ is basically considered as a \emph{duality} parameter, in the sense that the average values are computed by integrating $s$- and $-s$symbols of the observable and the density matrix; that is,}
\begin{equation}
\langle \hat{A}\rangle =\int \rmd\mu (\zeta ) \; W_{A}^{(s)}(\zeta )\,
W_{\varrho}^{(-s)}(\zeta ) = \int \rmd\mu (\zeta )\; 
W_{A}^{(-s)}(\zeta )\,W_{\varrho}^{(s)}(\zeta ) \, .  
\label{sW}
\end{equation}
\sugg{The Wigner function corresponds to $s=0$, so it is self-dual dual in this context. Unfortunately, the explicit construction of $s$-ordered maps and, especially, of the Wigner map is not as transparent as for the ${Q}$ and ${P}$ maps.}

When the group $G$ is compact, its \sugg{unitary representations are finite dimensional}  and the kernels $\hat{w}^{(s)}$ can be expanded in a basis of tensor operators $\{\hat{T}_{\nu}^{\lambda}\}$~\cite{Fano:1959ly} 
\begin{equation}
\hat{w}^{(s)}(\zeta ) = \sum_{\lambda ,\nu} w_{\lambda \nu}^{(s)}(\zeta ) \, 
\hat{T}_{\nu}^{\lambda} \, ,
\end{equation}
where $\lambda $ is a representation label appearing in the decomposition 
\begin{equation}
\Lambda \otimes \Lambda^{\ast} = \oplus n_{\lambda} \lambda \, ,
\label{eq:LambdaLambdastar}
\end{equation}
\sugg{where $n_{\lambda}$ is the number of times the irrep $\lambda$ appears in the decomposition and} the expansion coefficients $w_{\lambda \nu}^{(s)}(\zeta )$ can be expressed in terms of harmonic functions and appropriate Clebsch-Gordan coefficients~\cite{Brif:1999kx}.

When the Hilbert space of states is infinite-dimensional, delicate questions of convergence must be given careful attention, especially as the maps involve traces over infinitely many basis states of products of operators that can be formally represented by infinite-dimensional matrices. In particular, the decomposition of the product on the left hand side of (\ref{eq:LambdaLambdastar}) is non longer a direct sum but can include a direct integral of representations of the continuous type~\cite{Lindblad:1970aa,Repka:1978aa} making the construction of the irreducible tensor operators significantly more laborious and quite nontrivial~\cite{Holman:1966aa,Wang:1970aa}.

In the cases of locally flat classical phase-space corresponding to, e.g., the underlying $H(1)$ and $E(2)$ symmetries, sets of $s$-ordered map can be constructed ``by hand", in order to satisfy the basic requirements of normalization, invertibility and covariance under group action.

Except for the previous examples of noncompact symmetries and to the best of our knowledge, no self-dual maps from operators acting irreducibly in an infinite-dimensional Hilbert space into Wigner-like functions satisfying the Moyal-Stratanovich postulates have been discussed in details, even if applications of $SU(1,1)$ $Q$- and $P$- functions were discussed in \cite{Orowski:1990aa,Brif:1997aa,Kastrup:2003aa,Seyfarth:2020aa}.

In this paper we remedy this situation: we present practical expressions for the ${s}$-ordered Wigner functions of systems with ${SU}(1,1)$ symmetry using a connection between the ${Q}$ and ${P}$ maps through the action of an operator invariant under the group. Notably, a self-dual mapping kernel is obtained as a ``half-way" operator between $\hat{w}^{(+)}$ and $\hat{w}^{(-)} $~\cite{Figueroa:1990aa}. The phase-space functions are defined on the upper sheet of the two-sheet hyperboloid or equivalently in the interior of the Poincar\'{e} disc.

Beyond this solution to the technical problem of constructing ${SU}(1,1)$ Wigner functions, there are several reasons to investigate ${SU}(1,1)$ states in phase-space: ${SU}(1,1)$ plays a pivotal role in connection with what can be called two-photon effects~\cite{Wodkiewicz:1985aa,Gerry:1985aa,Gerry:1991aa,Gerry:1995kq}. The topic is experiencing a revival in popularity due to the recent realization of a nonlinear SU(1,1) interferometer~\cite{Jing:2011aa, Hudelist:2014aa}. According to the proposal of Yurke \textit{et al.}~\cite{Yurke:1986yg}, this device would allow one to improve the phase measurement sensitivity in a remarkable manner~\cite{Chekhova:2016aa,Li:2016aa}. In addition, the dynamics of such states strongly depends on the distinct possible plane sections of the hyperboloid \cite{Banerji:1999aa}.

\section{General setup for ${SU}(1,1)$}

\subsection{Coherent states and the coset space ${SU}(1,1)/U(1)$}

The Lie algebra $\mathfrak{su}(1,1)$ is spanned by the operators $\{\hat{K}_0,\hat{ K}_1,\hat{K}_2\}$ with commutation relations  
\begin{equation}
[\hat{K}_1,\hat{K}_2] = - \rmi \hat{K}_0\, , \qquad [ \hat{K}_2,\hat{K}_0] =
+ \rmi \hat{K}_1 \, , \qquad [ \hat{K}_0,\hat{K}_1] = + \rmi \hat{K}_2\, .
\end{equation}

We consider first a Hilbert space $\mathcal{H}$ that carries an irrep labelled by the Bargman index $k=\frac{1}{2},1,\frac{3}{2},2,\ldots $ of the group $G={SU}(1,1)$; the representation $k$ is in the positive discrete series. This explicitly excludes the single-mode even and odd harmonic oscillator states, which belong to the $k=\frac{1}{4}$ and $\frac{3}{4}$ irreps, respectively. 

\sugg{States in the irrep $k$ satisfy} 
\begin{equation}
\hat{K}_{0}|k,k+m\rangle = (k+m)|k,k+m\rangle \,, 
\qquad  
\hat{K}_{-}|k,k\rangle = 0 \,,   
\label{K0action}
\end{equation}
\sugg{where $m=0,1,\ldots$ and $\hat{K}_{\pm}=\pm \rmi(\hat{K}_{1}\pm \rmi\hat{K}_{2})$.} Let $H \subset G$ be the $U(1)$ subgroup of $G$ that leaves $|k,k\rangle $ invariant, up to a phase; $H$ is generated by exponentiating $\hat{K}_{0}$. The ${SU}(1,1)$ coherent states for the positive discrete series are labelled by points $\zeta $ in the interior of the Poincar\'{e} disc, $|\zeta |<1$, $\{|\zeta \rangle \in \mathcal{H},\zeta \in \mathcal{M=}{SU}(1,1)/U(1)\}$ and constructed as orbits of the cyclic vector $|k,k\rangle $~\cite{Perelomov:1986ly}, 
\begin{equation}
|\zeta \rangle = \hat{D}(\zeta )|k,k\rangle ,
\qquad 
\hat{D}(\zeta )= \rme^{\zeta \hat{K}_{+}} 
\rme^{-\ln (1-|\zeta |^{2})\hat{K}_{0}} 
\rme^{-\zeta ^{\ast}\hat{K}_{-}}\,.  
\label{CS}
\end{equation}
The unit disc can be lifted to the upper sheet of the two-sheeted hyperboloid by inverse stereographic map; this hyperboloid is our classical phase space, where points are parametrized by the hyperbolic Bloch vector 
\begin{equation}
\mathbf{n}=
(\cosh \tau ,\sinh \tau \cos \phi ,\sinh \tau \sin \phi )^{\top} \,,  
\label{BV}
\end{equation}
and where $\tau $ and $\phi $ are related to the complex number $\zeta $ through $\zeta =\tanh (\tau /2 ) \rme^{-\rmi\phi}$. 

\sugg{The symplectic 2-form on the hyperboloid~\cite{Perelomov:1986ly}}
\begin{equation}
\rmd \omega = \sinh \tau \, \rmd\tau \wedge \rmd\phi ,
\end{equation}
\sugg{induces the following Poisson bracket}
\begin{equation}
\{ f,g \} = \frac{1}{\sinh \tau} 
 \left( \frac{\partial f}{\partial \tau}
 \frac{\partial g}{\partial \varphi} - 
 \frac{\partial f}{\partial \varphi}
 \frac{\partial g}{\partial \tau}\right) \, ,  
 \label{PB}
\end{equation}
\sugg{where $f(\tau ,\phi )$  and $g(\tau ,\phi )$ are smooth functions. In particular, the components $\mathbf{n}=(n_{0},n_{1},n_{2})^{\top}$ of the Bloch vector (\ref{BV}) satisfy the  relations}
\begin{equation}
\{n_{1},n_{2}\} = - n_{0} \, ,
\qquad 
\{n_{2},n_{0}\} = n_{1} \, 
\qquad
\{n_{0}, n_{1}\} = n_{2} \, .
\end{equation}

In the basis $\{|k,k+m\rangle : m=0,1,\ldots\}$ the coherent states can be expanded as 
\begin{equation}
|\zeta \rangle = (1-|\zeta |^{2})^{k}\sum_{m=0}^{\infty} \left [ 
\frac{\Gamma(m+2k)}{m!\Gamma (2k)}\right ]^{1/2} 
\zeta^{m}|k,k+m\rangle \, ,
\label{cs11}
\end{equation}
and resolve the identity for $k>1/2$ 
\begin{equation}
\hat{\openone} = \frac{2k-1}{\pi}
\int \rmd \mu (\zeta )\, 
|\zeta \rangle \langle \zeta | \,,  
\label{Nk}
\end{equation}
(for $k=1/2$, the limit $k\rightarrow 1/2$ \sugg{must be taken in the final expressions}), where the invariant measure is given by 
\begin{equation}
\rmd \mu (\zeta ) = \frac{\rmd^{2}\zeta}{(1-|\zeta |^{2})^{2}} 
= \frac{1}{4} \sinh \tau \rmd\tau \rmd \phi , 
\qquad 
\rmd^{2}\zeta =\rmd  \re \zeta \; \rmd \im \zeta \, .  
\label{eq:invmeas}
\end{equation}
${SU}(1,1)$ coherent states are not orthogonal; their overlap in the discrete irrep $k$ is given by 
\begin{equation}
|\langle \zeta |\zeta^{\prime}\rangle |^{2} = 
\left ( \frac{1+\mathbf{n} \cdot \mathbf{n}^{\prime}}{2}\right)^{-2k} \, ,
\end{equation}
where $\mathbf{n}\cdot \mathbf{n}^{\prime}$ is a pseudo-scalar product on the hyperboloid, 
\begin{equation}
\mathbf{n} \cdot \mathbf{n}^{\prime}=\cosh \tau \cosh \tau^{\prime} 
-\cos (\phi -\phi^{\prime})\sinh \tau \sinh \tau^{\prime} 
\equiv \cosh \xi \, .
\label{nn'}
\end{equation}

\subsection{The kernels}

The $SU(1,1)$ quantization kernels $\hat{w}^{(s)}(\zeta )$, \sugg{generating dual maps according to (\ref{sW}), are operators} labelled by points of $\mathcal{M=}{SU}(1,1)/U(1)$. Their explicit form depends on the representation index $k$, but we will not explicitly write this dependence to avoid burdening the notation. The \emph{boundary} kernels $\hat{w}^{(\pm)}(\zeta )$ define direct and inverse projections on the set of coherent states (\ref{cs11})~\cite{Kastrup:2006cr}: 
\begin{eqnarray}
 \hat{A}=\frac{2k-1}{\pi}\int \rmd\mu (\zeta ){P}_{A}(\zeta )\,
 |\zeta \rangle \langle \zeta |\,,   \nonumber  \\
 \\
{P}_{A}(\zeta )=\Tr [ \hat{A} \hat{w}^{(+)}(\zeta )]\,,
\qquad \qquad 
{Q}_{A}(\zeta )=\Tr [ \hat{A} \hat{w}^{( -)}(\zeta )]\,,    \nonumber
\label{P}
\end{eqnarray}
and $\hat{w}^{(-)}(\zeta )=|\zeta \rangle \langle \zeta |$.

In~\ref{appA} we show that \sugg{there is a class of $s$-parametrized kernels} that are connected to $\hat{w}^{(\pm )}(\zeta )$ through the following relations: 
\begin{eqnarray}
\hat{w}^{(s)}(\zeta ) & = & \frac{2}{\pi}\int \rmd\mu (\zeta ^{\prime})
\int  \rmd\lambda \,\lambda \tanh (\pi \lambda )\,
\Phi_{k}^{\frac{1}{2}-\frac{s}{2}}(\lambda ) \, 
P_{-\frac{1}{2}+\rmi\lambda} (\zeta^{\prime -1} \zeta )
\hat{w}^{(+)}(\zeta ^{\prime})\,,  
\nonumber  \label{Rww+-} \\
& = & \frac{2}{\pi}\int \rmd(\zeta ^{\prime})\int \rmd\lambda \,
\lambda \tanh(\pi \lambda )\,\Phi_{k}^{-\frac{1}{2}-\frac{s}{2}}(\lambda )
\,P_{-\frac{1}{2}+\rmi\lambda}(\zeta ^{\prime -1} \zeta )
\hat{w}^{(-)}(\zeta ^{\prime}) \,, \nonumber \\
\end{eqnarray}
where $\Phi_{k}(\lambda )$ is 
\begin{equation}
\Phi_{k}(\lambda )=
\frac{(2k-1)|\Gamma (2k-\frac{1}{2}+\rmi\lambda )|^{2}}
{\Gamma ^{2}(2k)} \stackrel{\lambda \gg 1}{\sim}
\lambda^{4k-3/2} \rme^{-\pi \lambda} \, ,   
\label{Phi}
\end{equation}
and $P_{-\frac{1}{2}+i\lambda}(x)$ is the Legendre function~\cite{Erdelyi:1955aa,NIST:DLMF} with $P_{- \frac{1}{2}+\rmi\lambda}(\zeta^{\prime -1}\zeta )=P_{-\frac{1}{2}+\rmi\lambda}(\mathbf{n}\cdot \mathbf{n}^{\prime})$. \sugg{The invariant integration of the SU(1,1) covariant kernels $\hat{w}^{(\pm )}(\zeta )$ does warrant the covariance of the  family $\hat{w}^{(s)}(\zeta )$.}

By construction, the kernels (\ref{Rww+-}) satisfy the overlap relation 
\begin{equation}
\frac{2k-1}{4\pi}\Tr [ \hat{w}^{(s)}(\zeta ) 
\hat{w}^{(-s)}(\zeta^{\prime}) ] = 
\delta (\zeta ^{\prime},\zeta ) = 
\delta (\cosh \tau -\cosh \tau ^{\prime})\, 
\delta (\phi -\phi ^{\prime} ) \, ,  
\label{sd}
\end{equation}
and the normalization conditions 
\begin{equation}
\Tr [ \hat{w}^{(s)}(\zeta ) ]=1\,, 
\qquad \quad 
\frac{2k-1}{\pi}\int \rmd\mu(\zeta ) \, 
\hat{w}^{(s)}(\zeta )=\hat{\openone} \,.  
\label{ints}
\end{equation}
In particular, the Wigner symbol ($s=0$) of an operator $\hat{A}$ is related
to $Q$- and $P$- symbols by 
\begin{eqnarray}
\fl \qquad \qquad {W}_{A}(\zeta ) & \equiv & 
\Tr [ \hat{A}\hat{w}^{(0)}(\zeta )] \nonumber \\
&=&\frac{2}{\pi}\int \rmd\mu (\zeta ^{\prime})\,
{g}_{k}^{(+)}(\zeta^{\prime -1}\zeta ) 
{P}_{A}(\zeta ^{\prime}) = \frac{2}{\pi}
\int \rmd\mu(\zeta^{\prime}) 
{g}_{k}^{(-)}(\zeta ^{\prime -1}\zeta ){
Q}_{A}(\zeta^{\prime})\,,  
\label{W+-}
\end{eqnarray}
where 
\begin{equation}
{g}_{k}^{(\pm )}(\zeta ^{\prime -1}\zeta ) = 
\int_{0}^{\infty}\rmd\lambda \,\lambda \tanh (\pi \lambda )\;
\Phi_{k}^{\pm \frac{1}{2}}(\lambda )
P_{-\frac{1}{2}+\rmi\lambda}(\mathbf{n}\cdot \mathbf{n}^{\prime}) \, .  
\label{gk}
\end{equation}
In \sugg{consequence, the Wigner symbols satisfy the normalization} 
\begin{equation}
\frac{2k-1}{\pi}\int \rmd\mu (\zeta ){W}_{A}(\zeta )=1\,.
\end{equation}

The map (\ref{Ws}) generated by the kernels in (\ref{Rww+-}) is invertible in the standard sense: 
\begin{equation}
\hat{A} = \frac{2k-1}{\pi} \int \rmd \mu (\zeta ) \, 
W_{A}^{(s)}(\zeta )\, \hat{w}^{(-s)}(\zeta )\, .  
\label{inverse}
\end{equation}
The self-duality condition \sugg{of the Wigner map} is obviously satisfied here and average values are computed in accordance with equation~(\ref{sdW}): 
\begin{equation}
\langle \hat{A}\rangle = \frac{2k-1}{\pi} \int \rmd \mu (\zeta) 
W_{A} ( \zeta ) W_{\rho} ( \zeta ) \, .  
\label{averageW}
\end{equation}

We note that the equations~(\ref{Rww+-}) \sugg{can also be formally represented in the compact form} 
\begin{equation}
\hat{w}^{(s)}(\zeta ) = \Phi_{k}^{\frac{1}{2} - \frac{s}{2}}(\mathcal{L}^{2})
\, \hat{w}^{(+)}(\zeta ) = 
\Phi_{k}^{-\frac{1}{2}-\frac{s}{2}} (\mathcal{L}^{2})\, 
\hat{w}^{(-)}(\zeta ),  
\label{opW}
\end{equation}
with 
\begin{equation}
\Phi_{k}(\mathcal{L}^{2}) = - \frac{\pi \mathcal{L}^{2}}
{\cos ( \pi \sqrt{1/4+\mathcal{L}^{2}})} 
\prod_{m=1}^{2k-2} \left[ 1-\frac{\mathcal{L}^{2}}{m(m+1)}\right] \,,  
\label{PiopM}
\end{equation}
and {$\mathcal{L}^{2}$} is the Laplace operator on the hyperboloid~\cite{Alonso:2002aa} 
\begin{equation}  
\mathcal{L}^{2}=\frac{\partial^{2}}{\partial \tau^{2}}+ 
\coth \tau \frac{\partial}{\partial \tau} + 
\frac{1}{\sinh^{2}\tau}\frac{\partial^{2}}{\partial \varphi^{2}}.
\label{eq:Lphyp}
\end{equation}

The function ${g}_{k}^{(-)}$ \sugg{in equation~\ref{gk} is singular}, as one can see using the asymptotic behavior in (\ref{Phi}). This makes it inconvenient for calculations. In practice, the Wigner functions of physical states can be numerically generated only from the $P$-function; i.e., in terms of the ${g}_{k}^{(+)}$ function.

\sugg{It is worth noting that the relations~(\ref{Rww+-}) allow one to express the star product of $s$-parametrized symbols~\cite{Groenewold:1946aa}; i.e.,}
\begin{equation}
W_{fg}^{(s)}=W_{f}^{(s_{1})} \ast W_{g}^{(s_{2})} \, ,  
\label{sp}
\end{equation}
\sugg{in the integral form~\cite{Brif:1999kx}}
\begin{equation}
W_{fg}^{(s)} = \int \rmd \mu (\zeta_{1})\rmd \mu (\zeta_{2})
L_{s,s_{1},s_{2}} ( \zeta , \zeta_{1}, \zeta_{2} )
W_{f}^{(s_{1})}(\zeta_{1}) \, W_{g}^{(s_{2})}(\zeta_{2})  \, ,
\label{SP}
\end{equation}
\sugg{where}
\begin{equation}
L_{s,s_{1},s_{2}} ( \zeta , \zeta_{1}, \zeta_{2} ) = 
\Tr [ \hat{w}^{(s)}(\zeta ) \, \hat{w}^{(s_{1})}(\zeta_{1}) 
\, \hat{w}^{(s_{2})}(\zeta _{2}) ]\, .  
\label{K}
\end{equation}
\sugg{In particular, the Wigner symbol of a product of two operators can be conveniently represented in terms of the convolution of the corresponding $P$-symbols according to}
\begin{equation}
\fl 
W_{fg}^{(0)}=\Phi_{k}^{-\frac{1}{2}}(\mathcal{L}^{2})
\left( \frac{2k-1}{\pi}\right)^{2} 
\int \rmd \mu (\zeta_{1})\rmd \mu (\zeta_{2}) 
P_{f}(\zeta_{1}) \, P_{g}(\zeta_{2}) \,
\langle \zeta_{2}|\zeta \rangle \langle \zeta |\zeta_{1}\rangle 
\langle \zeta_{1}|\zeta_{2}\rangle \, .
\end{equation}

\section{Examples of Wigner functions}

\subsection{Coherent states}

\begin{figure}[b]
\centering
\subfigure[]{\includegraphics[width=70mm]{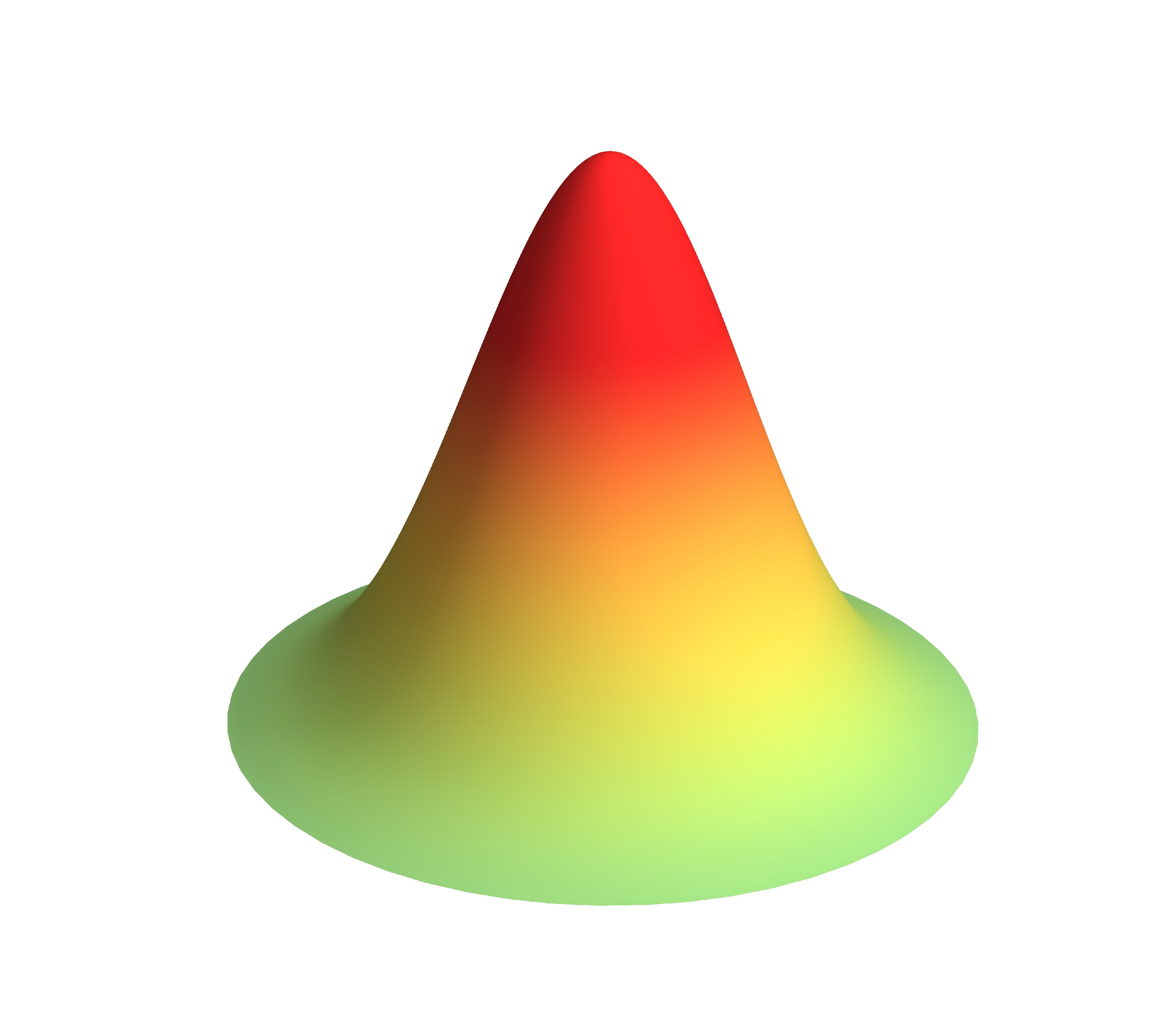}} \subfigure[]{%
\includegraphics[width=70mm]{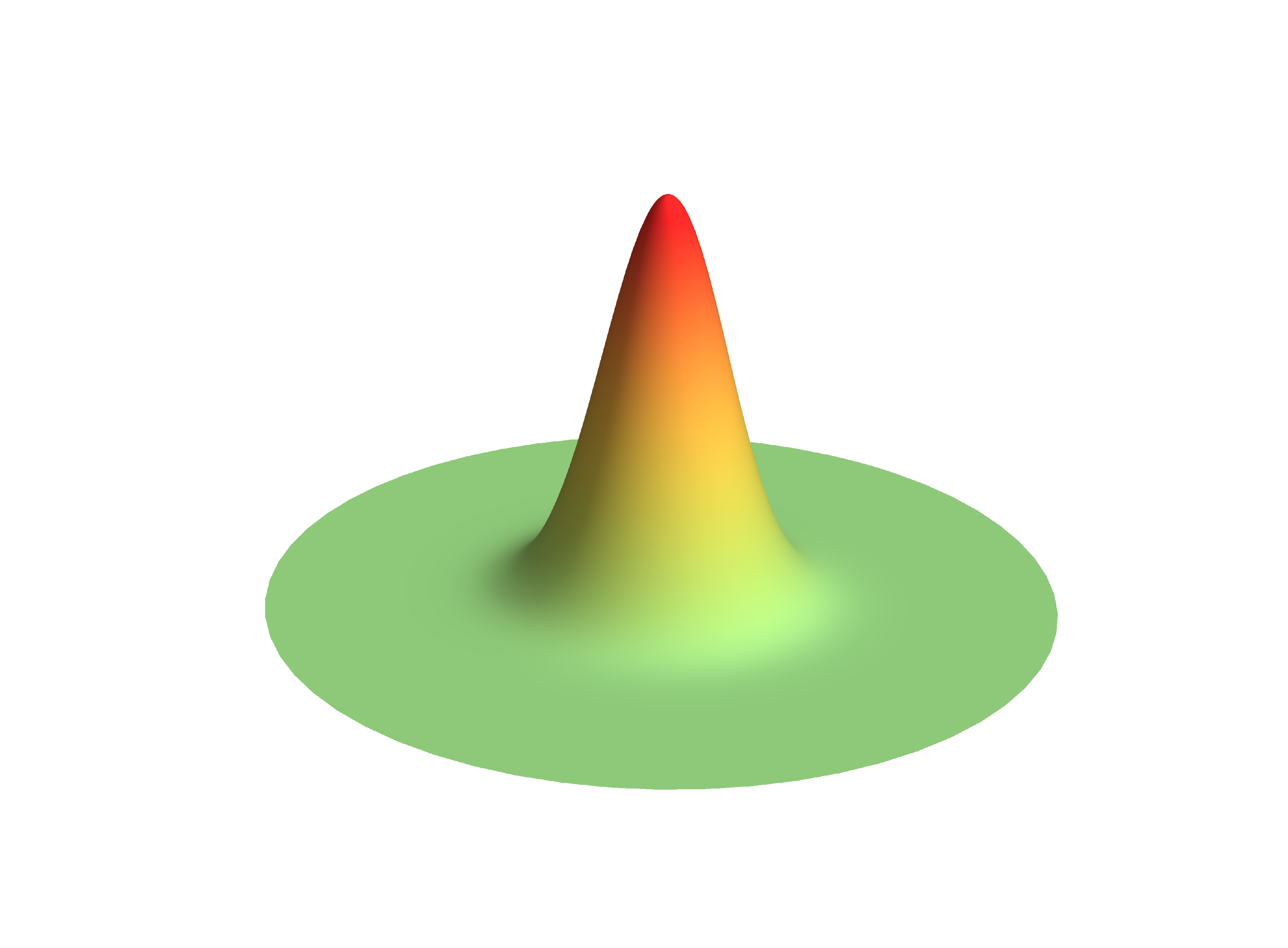}}
\caption{Plots of the SU(1,1) Wigner function of the ground state $%
|k,k\rangle$ on the Poincar\'{e} disc a) $k=1$; b) $k=5$.}
\label{fig:1}
\end{figure}

The Wigner function for $SU(1,1)$ coherent states is fairly easy to obtain using equation~(\ref{W+-}), since the ${P}$-function of a coherent state $|\zeta_{0}\rangle $, is a $\delta $-function on the hyperboloid: 
\begin{equation}
{P}_{|\zeta_{0}\rangle} (\zeta ) = \frac{4\pi}{2k-1} 
\delta (\zeta,\zeta_{0}) = \frac{4\pi}{2k-1} 
\delta (\cosh \tau -\cosh \tau_{0})\delta (\phi-\phi_{0}) \, .
\end{equation}
Then, the corresponding Wigner function is 
\begin{equation}
{W}_{|\zeta_{0}\rangle}(\zeta ) = \frac{2}{2k-1} 
{g}_{k}^{(+)}(\zeta_{0}^{-1}\zeta ) \, .
\end{equation}
In the particular case of the lowest weight state $|\zeta_{0}\rangle =|k,k\rangle $ the Wigner function is 
\begin{equation}
{W}_{|k,k\rangle} (\zeta ) = \frac{2}{2k-1} \int_{0}^{\infty} 
\rmd\lambda \; \lambda \tanh (\pi \lambda ) \, 
\Phi_{k}^{\frac{1}{2}}(\lambda ) \; 
P_{-\frac{1}{2}+ \rmi\lambda}(\cosh \tau )\,.  
\label{kkWF}
\end{equation}
In figure~\ref{fig:1} we plot the Wigner functions of equation~(\ref{kkWF}) of the ground state $|k,k\rangle $ as a distribution on the Poincar\'e disc for two irreps with $k=1$ and $k=5$ respectively. The distribution becomes narrower as $k$ increase. The difference in the scale is due to the normalization factor $\sim 2k-1$ appearing in~(\ref{ints}).

A more interesting case is the Wigner function for the superposition of two $SU(1,1)$ coherent states: 
\begin{equation}
|\Psi \rangle =\alpha | \zeta_{0}\rangle + \beta |\zeta_{1}\rangle \,.
\end{equation}
The corresponding Wigner functions exhibits interference and has the form
(see~\ref{appB}) 
\begin{equation}
{W}_{|\Psi \rangle}(\zeta ) = |\alpha |^{2} {W}_{|\zeta_{0}\rangle} (\zeta )
+ |\beta |^{2} {W}_{|\zeta_{1}\rangle}(\zeta )+  
2 \re [ \alpha \beta^{\ast} \, W_{\zeta_{0}\zeta_{1}}(\zeta ) ] \, ,
\end{equation}
where $W_{\zeta_{0}\zeta_{1}}(\zeta )$ is 
\begin{equation}
W_{\zeta_{0}\zeta_{1}}(\zeta )=\frac{2(1-|\zeta_{0}|^{2})^{k}
(1-|\zeta_{1}|^{2})^{k}}{(2k-1) (1- \zeta_{0}\zeta_{1}^{\ast})^{2k}}
{g}_{k}^{(+)} \left( \frac{2 ( 1-\zeta^{\ast} \zeta_{0}) 
(1-\zeta_{1}^{\ast} \zeta)}{(1-|\zeta |^{2})(1-\zeta_{0}
\zeta_{1}^{\ast})} - 1 \right) .
\end{equation}

\begin{figure}[t]
\centering
\subfigure[]{\includegraphics[width=70mm]{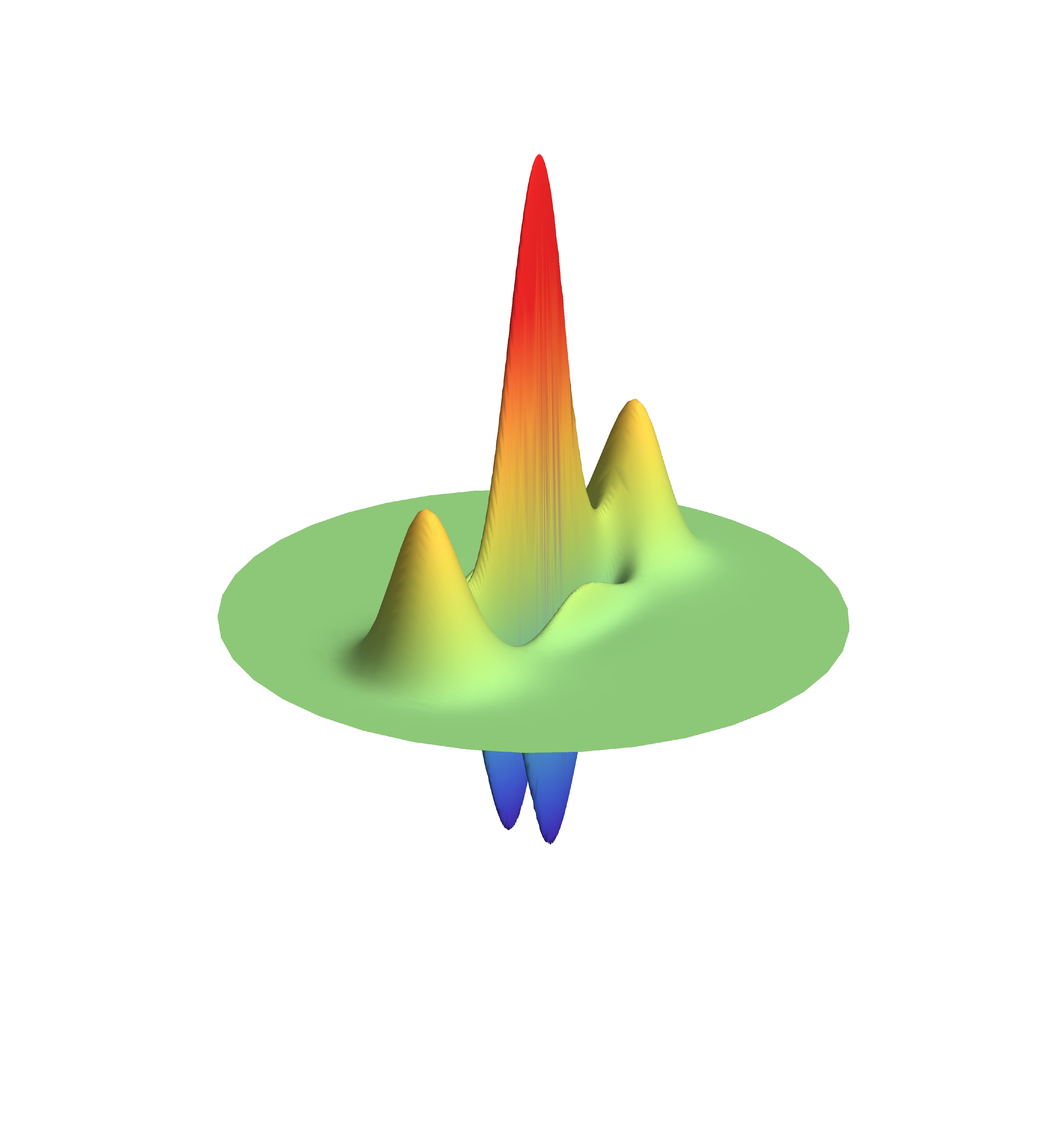}} \subfigure[]{%
\includegraphics[width=70mm]{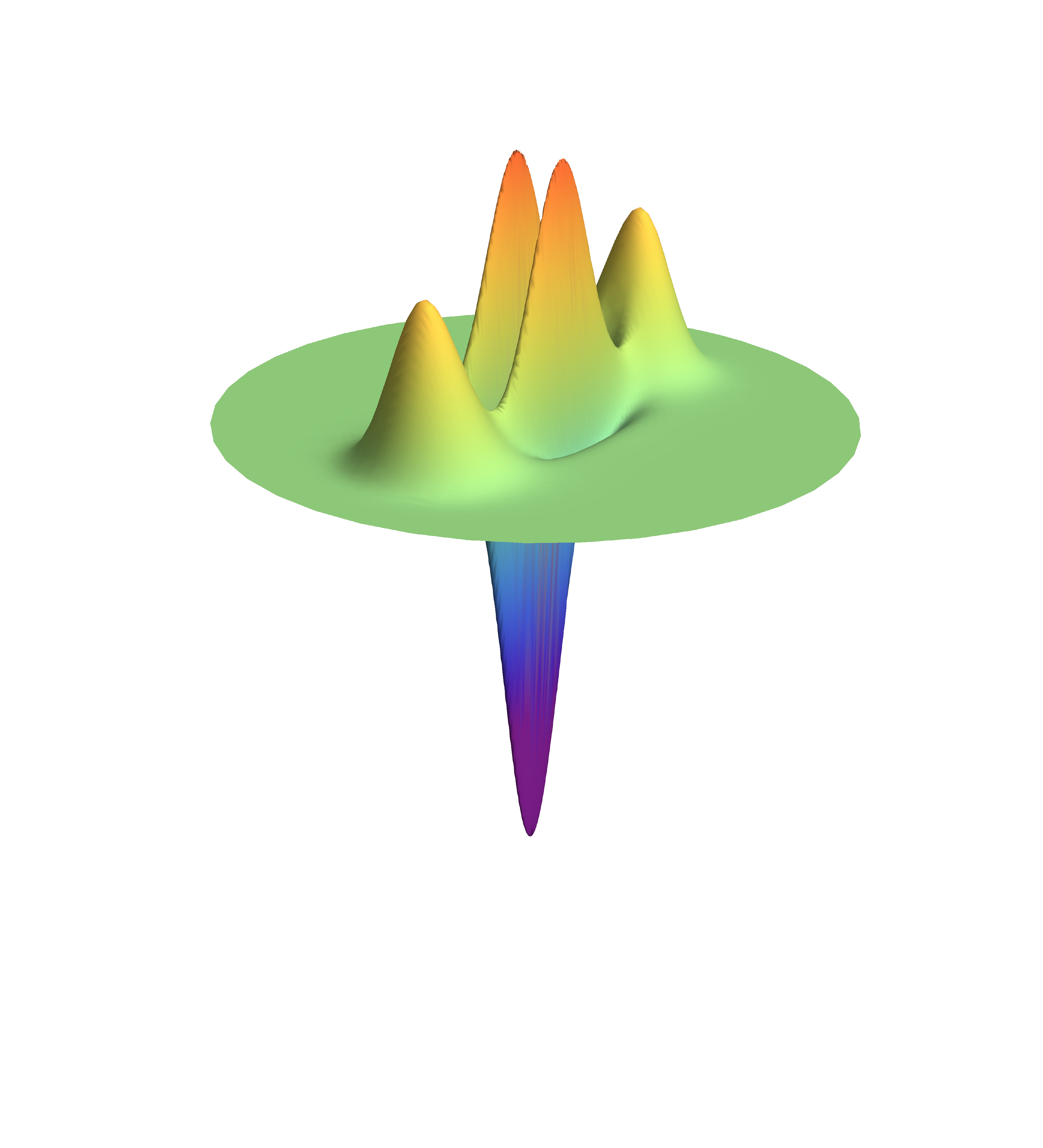}} 
\caption{Plots of the SU(1,1) Wigner function of the cat states in
equation~( \protect\ref{cats}) on the Poincar\'{e} disc a) even
superposition; b) odd superposition; in both cases $k=5$}
\label{fig:2}
\end{figure}

The Wigner function allows to visualize the interference pattern appearing in phase-space discription of pure states superposition, and thus distinguish them from mixed states. In figure~\ref{fig:2} we plot the Wigner function of even and odd superpositions of $SU(1,1)$ coherent states (cat-like states) 
\begin{equation}
|\Psi \rangle = \frac{N}{\sqrt{2}} 
(|\zeta_{0}\rangle \pm |-\zeta_{0}\rangle ) \, ,  
\label{cats}
\end{equation}
where $N= ( 1+\cosh^{-2k}\tau_{0} )^{-1/2}$.

The analytical expression for the Wigner function reads 
\begin{eqnarray}
{W}_{|\Psi \rangle}(\tau ,\phi ) & = & \frac{N^{2}}{2k-1} 
\int_{0}^{\infty}  \rmd \lambda \, \lambda \tanh ( \pi \lambda ) \, 
\Phi_{k}^{\frac{1}{2}}(\lambda ) \; \left[ 
P_{-\frac{1}{2}+\rmi\lambda} ( \cosh \xi_{+} ) \right. \nonumber \\
&+ &\left. P_{-\frac{1}{2}+\rmi \lambda} ( \cosh \xi_{-}) \pm 
\frac{2}{\cosh^{2k}\tau_{0}} \re P_{-\frac{1}{2}+\rmi \lambda}( z(\tau ,\phi )) 
\right] ,  \label{Wcats}
\end{eqnarray}
with 
\begin{eqnarray}
\cosh \xi_{\pm} & = & \cosh \tau \cosh \tau_{0} \mp \cos \phi \sinh \tau
\sinh \tau_{0} \, ,  \nonumber \\
&& \\
z(\tau ,\phi ) & = & \frac{\cosh \tau - \rmi \sinh \tau_{0}\sinh \tau \sin
\phi}{ \cosh \tau_{0}} \, .  \nonumber
\end{eqnarray}
The last term in equation~(\ref{Wcats}) describes the interference pattern. We point out that this pattern becomes more pronounced (i.e., the number of oscillatons increases) as the representation index $k$ grows. 

\subsection{Number states}

\begin{figure}[t]
\centering
\subfigure[]{\includegraphics[width=70mm]{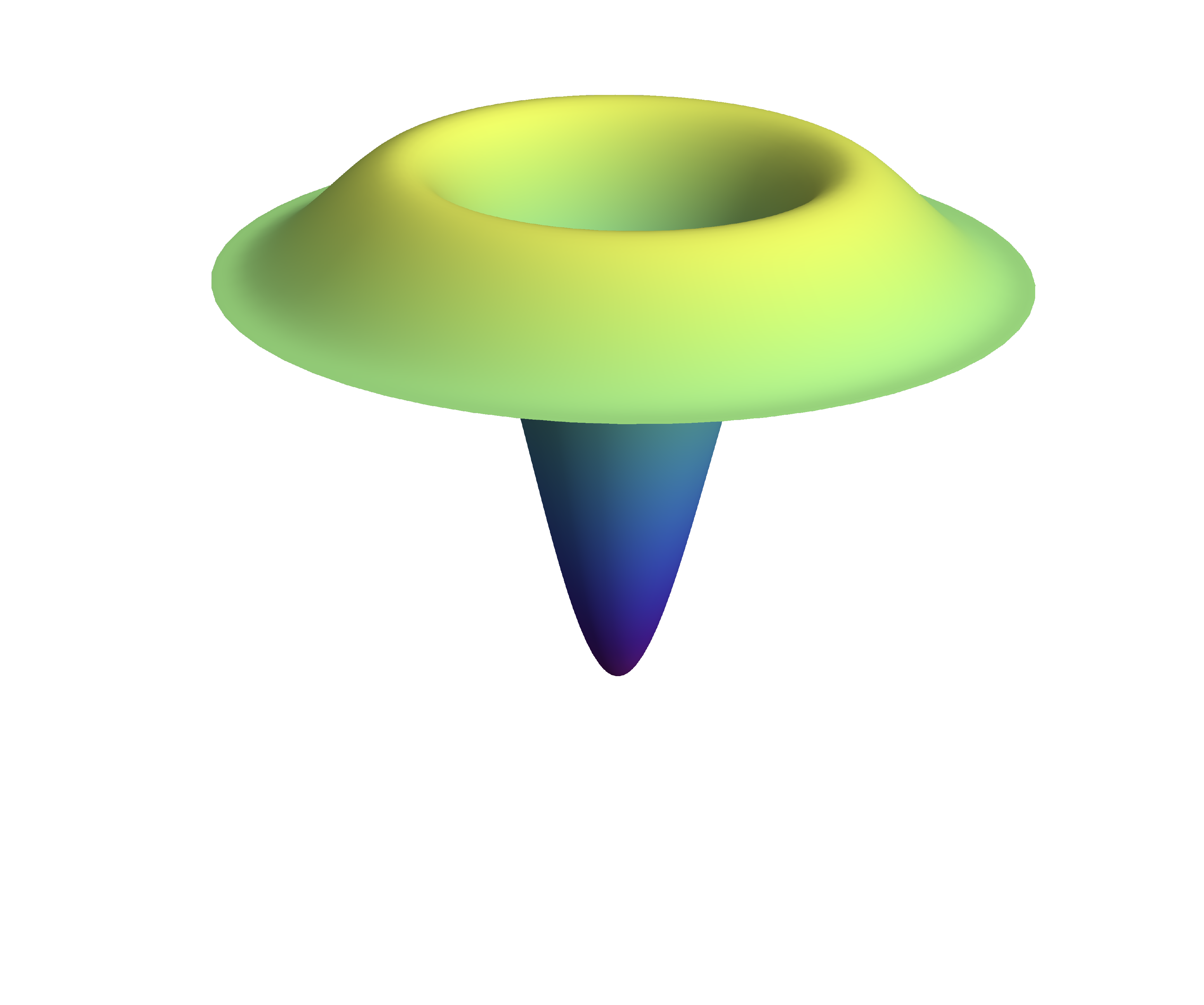}} 
\subfigure[]{\includegraphics[width=70mm]{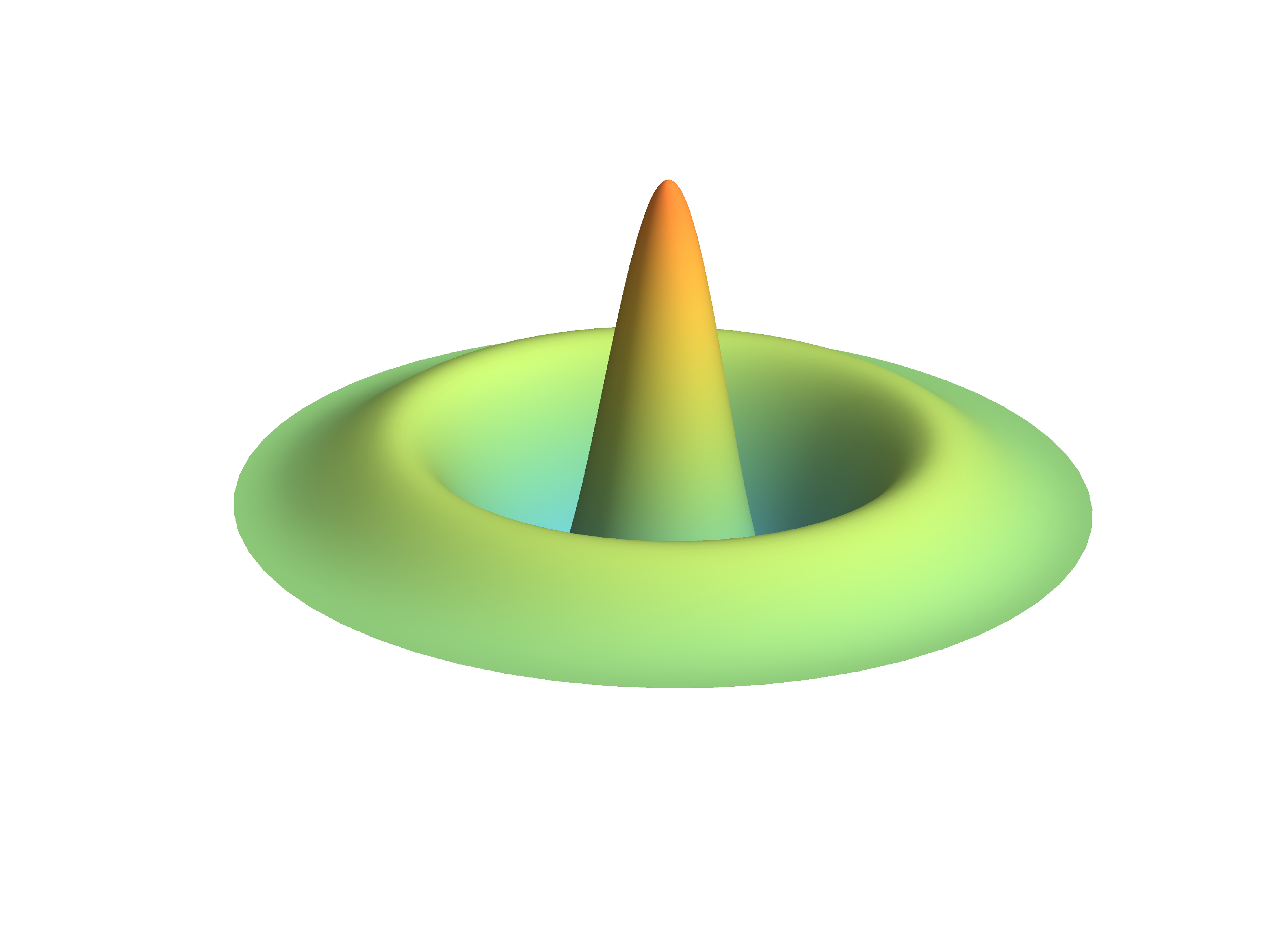}} 
\caption{Plots of the SU(1,1) Wigner function of the excited states on the
Poincar\'{e} disc a) $|k,k+1\rangle $ ; b) $|k,k+2\rangle $; in both cases $%
k=1$}
\label{fig:3}
\end{figure}

The Wigner function of the $SU(1,1)$ number states 
\begin{equation}
|k,k+m\rangle =\sqrt{\frac{\Gamma (2k)}{m!\Gamma (m+2k)}} \hat{K}
_{+}^{m}|k,k\rangle ,
\end{equation}
is obtained in~\ref{appB} and given by 
\begin{eqnarray}
\fl \qquad \qquad {W}_{|m\rangle}(\zeta ) & = & 
\frac{\Gamma (2k)} {(2k-1)\pi m! \Gamma (m+2k)} 
\int_{0}^{\infty} \rmd \lambda \, \lambda 
\tanh ( \pi \lambda ) \, \Phi_{k}^{\frac{1}{2}}(\lambda ) \nonumber   \\
& \times & \int \rmd \tau^{\prime} \rmd\phi^{\prime} 
\delta (\tau^{\prime}) [\cosh^{4} (\tau^{\prime}/{2}) \, 
{\mathcal{L}}^{\prime 2} ]^{m} [ \cosh^{4k} (\tau^{\prime}/2) \, 
P_{-\frac{1}{2}+\rmi \lambda}(\cosh \xi ) ] ,
\end{eqnarray}
where $\cosh \xi =\cosh \tau \cosh \tau^{\prime}-\cos (\phi-\phi^{\prime})
\sinh \tau \sinh \tau^{\prime}$ and where ${\mathcal{L}}^{\prime 2}$ is the Laplace operator in the hyperboloid, which acts on the primed variables.

The Wigner function of the first excited state is 
\begin{eqnarray}
 {W}_{|1\rangle}(\zeta ) &=&\frac{1}{( 2k-1) k} \int_{0}^{\infty} 
\rmd \lambda \, \lambda \tanh ( \pi \lambda ) \; 
\Phi_{k}^{\frac{1}{2}}(\lambda) (2k- 1/4 - \lambda^{2}) 
P_{-\frac{1}{2} + \rmi\lambda}( \cosh \tau )  \nonumber  \\
& = &\frac{1}{( 2k-1) k} \left( 2k + \frac{\partial^{2}}{\partial\tau^{2}}
+ \coth \tau \frac{\partial}{\partial \tau}\right) {g}_{k}^{(+)} ( \cosh
\tau ) \, .  \label{W1}
\end{eqnarray}
Figure~\ref{fig:3} illustrates the Wigner functions of the states $|k,k+1\rangle $ and $|k,k+3\rangle $ in the representation with $k=1$.

\section{Applications: $\mathfrak{su}(1,1)$ dynamics}

In quantum optics the $\mathfrak{su}(1,1)$ algebra naturally appears in the analysis of the non-degenerate parametric amplifier, with 
\begin{equation}
\hat{K}_{+}= \hat{a}^{\dagger} \hat{b}^{\dagger}, 
\qquad 
\hat{K}_{-} = \hat{a} \hat{b} , 
\qquad 
\hat{K}_{0} = \case{1}{2} ( \hat{a}^{\dagger} \hat{a} + 
\hat{b}^{\dagger}\hat{b} + \openone ) \, ,  
\label{tm}
\end{equation}
and where $\hat{a} $ and $\hat{b}$ are the standard boson operators. The coherent states (\ref{cs11}) form a convenient (but overcomplete) basis in each Hilbert space with a fixed difference $\Delta n$ of excitations between the modes $a$ and $b$. The $SU(1,1)$-irreducible subspaces are carrier spaces for irreps labelled by $k=\case{1}{2} ( 1+|\Delta n|) $. The evolution generated by Hamiltonians in the enveloping algebra of (\ref{tm}) can be suitably described as dynamics of $SU(1,1)$ quasidistributions on the hyperboloid or equivalently on the Poincar\'{e} disc.

The phase-space evolution on the hyperboloid generated by $\mathfrak{su}(1,1) $ Hamiltonians significantly differs from the dynamics on the two-dimensional sphere, the homogeneous space for $SU(2)$: while any Hamiltonian linear on the $SU(2)$ generators is equivalent to $\hat{H} =\omega \hat{S}_{z}$, there are compact and non-compact orbits in the case of the $SU(1,1)$ systems. In general, the dynamics of an initial state $
|\psi_{0}\rangle $ induced by an operator $T_{g}$ corresponding to a irrep of an element 
\begin{equation}
g=\left( 
\begin{array}{cc}
\alpha & \beta \\ 
\beta^{\ast} & \alpha^{\ast}
\end{array}
\right) ,\qquad 
|\alpha |^{2}-|\beta |^{2}=1 \, ,
\end{equation}
of the $SU(1,1)$ leads to an appropriate transformation of the Wigner function argument 
\begin{equation}
W_{T_{g}|\psi_{0}\rangle}(\zeta ) = W_{|\psi_{0}\rangle} \left ( 
\frac{-\alpha^{\ast}\zeta +\beta}{\beta^{\ast}\zeta -\alpha}\right) \, ,
\end{equation}
as a consequence of the Wigner function covariance under group transformations~\cite{Perelomov:1986ly}.

In particular, in case of compact evolution, the Hamiltonian 
\begin{equation}
\hat{H}=\chi \hat{K}_{0} \, ,  
\label{K0}
\end{equation}
generates rotation around the $z$-axis, and yields 
\begin{equation}
W_{|\zeta_{0}\rangle}(\zeta |t)= 
W_{|\zeta_{0}\rangle} ( \rme^{\rmi\chi t}\zeta ) \, ,
\end{equation}
or, equivalently, 
\begin{equation}
W_{|\zeta_{0}\rangle} (\tau ,\phi |t)= 
W_{|\zeta_{0}\rangle}(\tau,\phi-\chi t) \, .
\end{equation}
Any Hamiltonian $SU(1,1)$ equivalent to that in equation~(\ref{K0}) leads to a rotation of the initial distribution along an ellipse obtained as an intersection of the hyperboloid and an inclined plane.

The noncompact evolution is generated by $SU(1,1)$ Hamiltonians equivalent to 
\begin{equation}
\hat H=\chi \hat{K}_{2} \, .  
\label{Hn}
\end{equation}
For instance, the phase-space dynamics of the state $|\zeta_{0}=\tanh \tau_{0}/2\rangle $ governed by (\ref{Hn}) leads to 
\begin{equation}
W_{|\zeta_{0}\rangle}(\zeta |t) = 
W_{|\zeta_{0}\rangle}\left( 
\frac{\zeta\cosh \frac{\chi t}{2}+\sinh \frac{\chi t}{2}}
{\zeta \sinh \frac{\chi t}{2}+\cosh \frac{\chi t}{2}} \right) \, ,
\end{equation}
which explicitly exhibits a boost generated by (\ref{Hn}), e.g. 
\begin{equation}
W_{|\zeta_{0}\rangle}(\tau ,\phi =0|t) = 
W_{|\zeta_{0}\rangle}(\tau +\chi t,\phi =0) \, .
\end{equation}

\section{Concluding remarks}

In this work we have developed a basic and practical setup for a consistent  introduction of the Wigner map for the quantum systems with $SU(1,1)$ symmetry group acting irreducibly in a corresponding Hilbert space. The Wigner function generated by the kernels (\ref{W+-}) allow to faithfully represent states of quantum systems with underlying $SU(1,1)$ symmetry as distributions on the upper sheet of the hyperboloid or the Poincar\'{e} disc.

In the framework of our approach, the Wigner kernel can be formally obtained both from $Q$ and $P$ kernels. In a manner reminiscent of the Heisenberg-Weyl group, the transformation taking from $\hat{w}^{(-)}(\zeta )$ to $\hat{w}^{(0)}(\zeta )$ is singular. Thus, a practical way of obtaining the Wigner function is from the $P$-function of the corresponding state.

\ack
We dedicate this work to the memory of Prof. David J. Rowe, of the University of Toronto. The work of ABK is partially supported by the Grant 254127 of CONACyT (Mexico); HdG is supported in part by NSERC of Canada, LLSS is supported by the Spanish Ministerio de Ciencia e Innovaci\'on (Grant PGC2018- 099183-B-I00).

\appendix

\section{Properties of $\hat w^{(s)}$}

\label{appA}

We start with a full set of Perelomov-type coherent states $\{|\zeta \rangle \in \mathcal{H}\}$ generated from a fiducial state $|\psi_{0}\rangle $ and labelled by coordinates $\zeta $ of $\mathcal{M}$, a homogeneous space of \sugg{the} dynamical symmetry group $G=SU(1,1)$. We further assume that $\mathcal{H}$ carries an irrep $\Lambda $ in the positive discrete series of $SU(1,1)$,  \sugg{labelled by the Bargman index}$k=\frac{1}{2},1,\frac{3}{2},2,\ldots.$ Here, $\mathcal{M}=SU(1,1)/U(1)$ where $U(1)$ is the subgroup generated by $\hat{K}_{0}$.

The ${Q}$-and ${P}$-kernels $\hat{w}^{\left( \pm \right)}(\zeta )$, are connected through the relation 
\begin{equation}
\hat{w}^{(-)}(\zeta ) = \frac{2k-1}{\pi} \int \rmd\mu (\zeta^{\prime}) 
| \langle \zeta^{\prime}|\zeta \rangle |^{2} \; 
\hat{w}^{(+)}(\zeta^{\prime}) \, ,  
\label{wwsu11}
\end{equation}
where $\rmd\mu (\zeta )$ is the invariant measure (\ref{eq:invmeas}). They satisfy the duality relation 
\begin{equation}
\frac{2k-1}{4\pi}\, 
\Tr [ \hat{w}^{(+)}(\zeta^{\prime}) \hat{w}^{(-)}(\zeta) ] = 
\delta (\zeta ,\zeta^{\prime}) = 
\delta (\cosh \tau^{\prime}-\cosh \tau )\delta (\phi^{\prime}-\phi )\, .  
\label{dual}
\end{equation}

Following the general ideas of \cite{Figueroa:1990aa} we observe that 
\begin{equation}
\delta (\cosh \tau -\cosh \tau^{\prime})\delta (\phi -\phi^{\prime})= 
\frac{1}{2\pi}\sum_{n=-\infty}^{\infty} \int \rmd\lambda \; \lambda 
\tanh (\pi\lambda) \, u_{n}^{\lambda}(\zeta ) 
u_{n}^{\lambda \ast} (\zeta^{\prime}) \; ,
\end{equation}
where 
\begin{eqnarray}
u_{n}^{\lambda}(\zeta ) & = & \frac{1}{2\pi}\int_{0}^{2\pi}\rmd \theta \, 
[ \cosh \tau -\sinh \tau \cos (\theta -\phi )]^{-\frac{1}{2}+\rmi\lambda} 
\rme^{ \rmi n \theta}   \nonumber \\
& = & ( -1 )^{n}\frac{\Gamma (\frac{1}{2}+\rmi \lambda )}
{ \Gamma (\frac{1}{2}+\rmi\lambda +n)} \; 
P_{-\frac{1}{2}+\rmi\lambda}^{n} (\cosh \tau ) \rme^{\rmi n \phi} \, ,  
\label{un}
\end{eqnarray}
are the harmonic functions on the upper sheet of the hyperboloid $\mathcal{M} ={SU}(1,1)/U(1)$. The functions $u_{n}^{\lambda}(\zeta )$ are eigenfunctions of the Laplace operator {$\mathcal{L}^{2}$} (\ref{eq:Lphyp}) on the hyperboloid 
\begin{equation}
{\mathcal{L}^{2}}u_{n}^{\lambda}(\zeta ) = 
-\left( \lambda^{2}+\frac{1}{4}\right) u_{n}^{\lambda}(\zeta )\,,  
\label{casimir11}
\end{equation}
and satisfy the following sum rule~\cite{Erdelyi:1955aa}, defining the zonal functions on ${SU}(1,1)/U(1)$: 
\begin{equation}
\sum_{n=-\infty}^{\infty}u_{n}^{\lambda}(\zeta ) 
u_{n}^{\ast\lambda}(\zeta^{\prime}) = 
P_{-\frac{1}{2}+\rmi \lambda} (\cosh \xi ) \, ,
\label{Phf}
\end{equation}
and $\cosh \xi$ has been defined in~(\ref{nn'}).

The harmonic functions of equation~(\ref{un}) also satisfy the orthogonality condition 
\begin{equation}
\lambda \tanh (\pi \lambda )\int \rmd\tau \rmd\phi \,\sinh \tau
\,u_{n}^{\lambda}(\zeta )u_{n^{\prime}}^{\lambda ^{\prime}\ast}(\zeta ) 
=2\pi \delta_{nn^{\prime}}\delta (\lambda -\lambda ^{\prime}).
\end{equation}
The expansion of a function $f(\zeta )$ on a hyperboloid on the basis of $ u_{n}^{\lambda}(\zeta )$ has thus the form  
\begin{equation}
\fl 
f(\zeta ) = \sum_{n=-\infty}^{\infty}\int \rmd\lambda \lambda 
\tanh (\pi \lambda ) \,u_{n}^{\lambda}(\zeta )f_{n\lambda}\, ,
\qquad 
f_{n\lambda} = \int \rmd\mu (\zeta )\,
u_{n}^{\lambda \ast}(\zeta )f(\zeta ) \, .  
\end{equation}
The functions $u_{n}^{\lambda}(\zeta )$ are nothing but the representation of elements of the basis of the principal continuous series, labelled by $-\frac{1}{2}+\rmi\lambda $,~\cite{Perelomov:1986ly} 
\begin{equation}
\hat{K}_{0}|\lambda ,n\rangle =  n|\lambda ,n\rangle \,, 
\qquad 
\hat{K}_{\pm}|\lambda ,n\rangle = \left( \pm \case{1}{2}\mp 
\rmi\lambda +n\right) |\lambda ,n\rangle \,,   
\end{equation}
\sugg{with  $n \in \mathbb{Z}$ and} $u_{n}^{\lambda}(\zeta )=\langle \zeta |\lambda ,n\rangle $.

It is easy to see that a differential operator $\hat{\Phi}_{\Lambda}(\zeta ) $, depending explicitly \sugg{on the Bargman index $k$ that labels} the representation $\Lambda $ and returning the squared coherent state overlap $|\langle \zeta ^{\prime}|\zeta \rangle |^{2}$ from $ \delta (\zeta ^{\prime},\zeta )$ should be invariant under group transformations: given $\hat{\Phi}_{\Lambda}(\zeta )\delta (\zeta ^{\prime},\zeta )=|\langle \zeta ^{\prime}|\zeta \rangle |^{2}$, then, by transitivity of $|\langle \zeta ^{\prime}|\zeta \rangle |^{2}$ and $\delta (\zeta ,\zeta ^{\prime})$ we have 
\begin{equation}
\fl\hat{\Phi}_{\Lambda}(g\zeta )\delta (g\zeta ^{\prime},\zeta )=|\langle
g\zeta ^{\prime}|\zeta \rangle |^{2}=|\langle \zeta ^{\prime}|g^{-1}\zeta
\rangle |^{2}=\hat{\Phi}_{\Lambda}(\zeta )\delta (\zeta ^{\prime
},g^{-1}\zeta )=\hat{\Phi}_{\Lambda}(\zeta )\delta (g\zeta ^{\prime},\zeta
),
\end{equation}
where $g\in SU(1,1)$. Thus, the operator $\hat{\Phi}_{\Lambda}
(\zeta )\equiv \hat{\Phi}_{k} (\zeta )$ is conveniently
expressed as a \emph{function} $\Phi_{k}$ of the operator $\mathcal{L}^{2}$
, the differential realization of the quadratic Casimir $\mathcal{C}_{2}$ on
the hyperboloid: 
\begin{equation}
\hat{\Phi}_{k}(\zeta )=\Phi_{k}(\mathcal{L}^{2}).
\end{equation}
Explicitly, for the square of the scalar product of two $SU(1,1)$ coherent
states in the representation labelled with $k=1/2,1,3/2,...$ we have 
\begin{eqnarray}
\fl\frac{2k-1}{4\pi}|\langle \zeta ^{\prime}|\zeta \rangle |^{2} &=&\frac{
2k-1}{4\pi}\left( \frac{1+\cosh \xi}{2}\right) ^{-2k}=\hat{\Phi}_{k}({
\mathcal{L}^{2}})\delta (\cosh \tau -\cosh \tau ^{\prime})\delta (\phi
-\phi ^{\prime})  \nonumber  \\
\;\; &=&\frac{1}{2\pi}\int \rmd\lambda \;\lambda \tanh \left( \pi \lambda
\right) P_{-\frac{1}{2}+\rmi\lambda}(\cosh \xi )\Phi_{k}(\lambda ).
\label{Phik}
\end{eqnarray}
In consequence, equation~(\ref{wwsu11}) can be rewritten as 
\begin{equation}
\hat{w}^{(-)}(\zeta )=\frac{2}{\pi}\int \rmd\mu (\zeta ^{\prime})\hat{w}
^{(+)}(\zeta ^{\prime})\int \rmd\lambda \;\lambda \tanh (\pi \lambda )P_{-
\frac{1}{2}+\rmi\lambda}(\cosh \xi )\Phi_{k}(\lambda )\,.  \label{w-w+}
\end{equation}

The inversion of equation~(\ref{Phik}) is given by~\cite{Erdelyi:1955aa} 
\begin{equation}
\Phi_{k}(\lambda )=\frac{2k-1}{2}\int_{1}^{\infty}\rmd x \left( \frac{1+x}{2
}\right)^{-2k} P_{-\frac{1}{2}+\rmi \lambda}(x).
\end{equation}
The above integral can be exactly computed with the result 
\begin{equation}
\Phi_{k}(\lambda )=\frac{\left( 2k-1\right) |\Gamma \left( 2k-\frac{1}{2}+ 
\rmi\lambda \right) |^{2}}{\Gamma^{2}(2k)},
\end{equation}
and its normalization follows from equation~(\ref{Phik}) 
\begin{equation}
\frac{2}{2k-1}\int \rmd\lambda \; \lambda \tanh ( \pi \lambda) \Phi
_{k}(\lambda )=1.
\end{equation}

Formally, one can represent equation~(\ref{w-w+}) in an operational form 
\begin{equation}
\hat{w}^{(-)}(\zeta )=\Phi_{k}(\mathcal{L}^{2})\hat{w}^{(+)}(\zeta ),
\label{opw}
\end{equation}
where $\Phi_{k}(\mathcal{L}^{2})$ is given in equation~(\ref{PiopM}). Now, 
\sugg{we can formally introduce ${s}$-parametrized kernels $\hat{w}^{(s)}(\zeta )$  related to $\hat{w}^{(\pm )}(\zeta )$ as}
\begin{eqnarray}
\hat{w}^{(s)}(\zeta ) &=&\frac{2}{\pi}\int \rmd\mu (\zeta ^{\prime})\hat{w}
^{(+)}(\zeta ^{\prime})\int \rmd\lambda \;\lambda \tanh (\pi \lambda )P_{-
\frac{1}{2}+\rmi\lambda}(\cosh \xi )\Phi_{k}^{\frac{1}{2}-\frac{s}{2}
}(\lambda )\, \nonumber \\
&= &\frac{2}{\pi}\int \rmd\mu (\zeta ^{\prime})\hat{w}^{(-)}(\zeta ^{\prime
})\int \rmd\lambda \;\lambda \tanh (\pi \lambda )P_{-\frac{1}{2}+\rmi\lambda
}(\cosh \xi )\Phi_{k}^{-\frac{1}{2}-\frac{s}{2}}(\lambda ) \nonumber \\
\end{eqnarray}
\sugg{that satisfy the overlap relation} 
\begin{equation}
\frac{2k-1}{4\pi}\Tr [ \hat{w}^{(s)}(\zeta ) \hat{w}^{(-s)}(\zeta^{\prime})
]=\delta (\zeta ^{\prime},\zeta )=\delta (\cosh \tau -\cosh \tau ^{\prime
})\delta (\phi -\phi ^{\prime \prime}).
\end{equation}
In particular, the self-dual Wigner kernel, $s=0$, is obtained from $\hat{w}
^{(\pm )}(\zeta )$ kernels by 
\begin{eqnarray}
\hat{w}^{(0)}(\zeta ) &=&\frac{2}{\pi}\int \rmd\lambda \;\lambda \tanh (\pi
\lambda )\Phi_{k}^{1/2}(\lambda )\int \rmd\mu (\zeta ^{\prime})\hat{w}
^{(+)}(\zeta ^{\prime})P_{-\frac{1}{2}+\rmi\lambda}(\cosh \xi ) \nonumber  \\
&=&\Phi_{k}^{1/2}(\mathcal{L}^{2})\hat{w}^{(+)}(\zeta )\, , \nonumber   \label{w+} \\
&& \\
\hat{w}^{(0)}(\zeta ) &=&\frac{2}{\pi}\int \rmd\lambda \;\lambda \tanh (\pi
\lambda )\Phi_{k}^{-1/2}(\lambda )\int \rmd\mu (\zeta ^{\prime})\hat{w}
^{(-)}(\zeta ^{\prime})P_{-\frac{1}{2}+\rmi\lambda}(\cosh \xi ) \nonumber  \\
&=&\Phi_{k}^{-1/2}(\mathcal{L}^{2})\hat{w}^{(-)}(\zeta ) \, . \nonumber  
\label{w-}
\end{eqnarray}
In this way, $\hat{w}^{(0)}(\zeta )$ automatically satisfies the
self-duality condition 
\begin{equation}
\frac{2k-1}{4\pi}\Tr [ \hat{w}^{(0)}(\zeta )\hat{w}^{(0)}(\zeta
^{\prime}) ]=\delta (\cosh \tau -\cosh \tau ^{\prime})\delta (\phi
-\phi ^{\prime}) \, .
\end{equation}
Since the kernels $\hat{w}^{(\pm )}(\zeta )$ satisfy the normalization
conditions~(\ref{ints}), one obtains from equation~(\ref{w-}) 
\begin{equation}
\Tr [ \hat{w}^{(0)}(\zeta ) ]=\Phi_{k}^{1/2}(\mathcal{L}^{2})\Tr [
\hat{w}^{(+)}(\zeta )]=1,
\end{equation}
since $\Phi_{k}(\mathcal{L}^{2})\,1=1$. In addition, using the
self-adjoitness of $\Phi_{k}(\mathcal{L}^{2})$ one has 
\begin{eqnarray}
\fl\frac{2k-1}{\pi}\int \rmd\mu (\zeta )\hat{w}^{(0)}(\zeta ) &=&\frac{2k-1
}{\pi}\int \rmd\mu (\zeta )\Phi_{k}^{1/2}(\mathcal{L}^{2})\hat{w}
^{(+)}(\zeta )=\frac{2k-1}{\pi}\int \rmd\mu (\zeta )\hat{w}^{(+)}(\zeta )=
\hat{\openone} \,. \nonumber  \\
&&
\end{eqnarray}
It is straightforward to obtain the average of the Wigner kernel over the
coherent states; i.e., the $Q$-function of the Wigner kernel 
\begin{equation}
\langle \zeta ^{\prime}|\hat{w}^{(0)}(\zeta )|\zeta ^{\prime}\rangle =
\frac{2}{2k-1}\int \rmd\lambda \,\lambda \tanh (\pi \lambda )P_{-\frac{1}{2}+
\rmi\lambda}(\cosh \xi )\Phi_{k}^{1/2}(\lambda ),
\end{equation}
which is a convergent integral.

\section{Wigner functions of some number states and superpositions}

\label{appB}

In this Appendix we obtain the Wigner functions of the number states and
nondiagonal projector on the coherent states. In order to obtain the Wigner
function of the $SU(1,1)$ number states 
\begin{equation}
|k,k+m\rangle =\sqrt{\frac{\Gamma (2k)}{m!\Gamma (m+2k)}} \hat{K}
_{+}^{m}|k,k\rangle ,  \label{m}
\end{equation}
we notice that 
\begin{eqnarray}
\fl \qquad \qquad \hat{K}_{+}^{m}|k,k\rangle \langle k,k|\hat{K}_{-}^{n} & =
& \frac{2k-1}{\pi} \int \hat{K}_{+}^{m}|\zeta \rangle \langle \zeta |\hat{K}
_{-}^{n}{P}_{|k,k\rangle}(\zeta )  \nonumber  \\
& = & \frac{2k-1}{\pi}\int \rmd\mu (\zeta )\left[ D_{L}^{m}(\hat{K}
_{+})D_{R}^{n}(\hat{K}_{-})|\zeta \rangle \langle \zeta |\right] {P}
_{|k,k\rangle}(\zeta ),
\end{eqnarray}
where 
\begin{equation}
\fl D_{L}(\hat{K}_{+}) = (1-|\zeta |^{2})^{2k}\partial_{\zeta}
(1-|\zeta|^{2})^{-2k} \, , \qquad D_{R}(\hat{K}_{-}) = (1-|\zeta
|^{2})^{2k}\partial_{\zeta^{\ast}}(1-|\zeta |^{2})^{-2k},
\end{equation}
and 
\begin{equation}
{P}_{|k,k\rangle}(\zeta )=\frac{2}{2k-1}\frac{1}{\sinh \tau} \delta (\tau )
\end{equation}
is the ${P}$-symbol for the lowest weight state $|k,k\rangle $ of irrep $k$.

In consequence, the ${P}$-function corresponding to the matrix element $
|k,k+m\rangle \langle k,k+n|$ has the form 
\begin{eqnarray}
{P}_{mn}(\zeta )& =\frac{(-1)^{m+n}}{(1-|\zeta |^{2})^{2k-2}}{N}_{k;mn} \;
\partial_{\zeta}^{m}\partial_{\zeta^{\ast}}^{n} [ (1-|\zeta|^{2})^{2k-2} \,{P
}_{|k,k\rangle}(\zeta ) ] ,  \nonumber \label{Pnm} \\
& & \\
{N}_{k;mn}& =\frac{\Gamma (2k)}{\sqrt{m!n!\Gamma (m+2k)\Gamma (n+2k)}}\,. \nonumber
\end{eqnarray}
Substituting the above expression into equation~(\ref{W+-}) and integrating
by parts we obtain after simplification the Wigner symbol of $|k,k+m\rangle
\langle k,k+n|$, 
\begin{eqnarray}
{W}_{mn}(\zeta ) & = & \frac{{N}_{k;mn}}{\left( 2k-1\right) \pi}
\int_{0}^{\infty}\rmd\lambda \; \lambda \tanh ( \pi \lambda ) \Phi_{k}^{
\frac{1}{2}}(\lambda ) \nonumber \\
& \times & \int \rmd\tau^{\prime}\rmd\phi^{\prime}\delta
(\tau^{\prime})\partial_{\zeta^{\prime}}^{m}\partial_{\zeta ^{\prime
\ast}}^{n}\left[ \cosh^{4k} (\tau^{\prime}/{2}) \, P_{-\frac{1}{2}+\rmi
\lambda}(\cosh \xi )\right] ,  \label{Wmn}
\end{eqnarray}
where 
\begin{eqnarray}
\partial_{\zeta}& =\mathrm{e}^{\rmi \phi}\cosh^{2} ( \tau/2 ) \partial
_{\tau} + \frac{\rmi}{2} \rme^{\rmi\phi} \coth ( \tau / 2 ) \partial_{\phi}
\, ,  \nonumber \\
& & \\
\partial_{\zeta}& =\mathrm{e}^{- \rmi \phi}\cosh^{2} ( \tau/2 ) \partial
_{\tau} - \frac{\rmi}{2} \rme^{- \rmi\phi} \coth ( \tau / 2 )
\partial_{\phi} \,  . \nonumber 
\end{eqnarray}
The Wigner function of the state (\ref{m}) is immediatly obtained from (\ref{Wmn}).

In order to compute the symbol $W_{\zeta_{0}\zeta_{1}}(\zeta )$ of the
nondiagonal projector $|\zeta_{0}\rangle \langle \zeta_{1}|$ we note that 
\begin{eqnarray}
\fl \qquad |\zeta_{0}\rangle \langle \zeta_{1}| &= &(1-|\zeta
_{0}|^{2})^{k} (1-|\zeta_{1}|^{2})^{k}  \nonumber \\
&\times & \sum_{m,n=0}^{\infty}\left[ \frac{\Gamma (m+2k)}{m!\Gamma (2k)}
\right ]^{1/2} \left[ \frac{\Gamma (n+2k)}{n!\Gamma (2k)}\right ]^{1/2}
\zeta_{0}^{m}\zeta_{1}^{\ast n}|k,k+m\rangle \langle k,k+n|.   
\end{eqnarray}
Recalling that the $P$-symbol of the matrix element $|k,k+m\rangle \langle
k,k+n|$ is given in equation~(\ref{Pnm}), we obtain the $P$-symbol of $
|\zeta_{0}\rangle \langle \zeta_{1}|$: 
\begin{eqnarray}
P_{\zeta_{0}\zeta_{1}}(\zeta ) & = & (1-|\zeta_{0}|^{2})^{k}
(1-|\zeta_{1}|^{2})^{k}(1-|\zeta |^{2})^{-2k+2}   \nonumber \\
&\times&\exp (-\zeta_{0}\partial_{\zeta}- \zeta_{1}^{\ast}\partial
_{\zeta^{\ast}}) [ (1-|\zeta |^{2})^{2k-2}P_{|k,k\rangle}(\zeta )] \, .
\end{eqnarray}
Substituting the above into equation~(\ref{W+-}) and integrating by parts
yelds 
\begin{eqnarray}
\fl \qquad W_{\zeta_{0}\zeta_{1}}(\zeta ) & = & \frac{4}{( 2k-1 ) \pi}
\int_{0}^{\infty}\rmd \lambda \, \lambda \tanh ( \pi \lambda )\; \Phi_{k}^{
\frac{1}{2}}(\lambda )(1-|\zeta_{0}|^{2})^{k}(1-|\zeta_{1}|^{2})^{k} \nonumber 
  \label{ksiksi} \\
& \times & \int \rmd\mu (\zeta^{\prime})\frac{\delta (\tau^{\prime})}{\sinh
\tau^{\prime}}\exp (\zeta_{0}\partial_{\zeta^{\prime}}+
\zeta_{1}^{\ast}\partial_{\zeta^{\prime \ast}}) [ (1-|\zeta^{\prime
}|^{2})^{-2k}P_{-\frac{1}{2}+\rmi\lambda}(\cosh \xi ) ] ,   
\end{eqnarray}
where now 
\begin{equation}
\cosh \xi =\frac{2|1-\zeta^{\ast}\zeta^{\prime}|^{2}}{(1-|\zeta
|^{2})(1-|\zeta^{\prime}|^{2})}-1 \, .
\end{equation}
Integrating equation~(\ref{ksiksi}) over $\mu (\zeta^{\prime})$ yields 
\begin{eqnarray}
\fl W_{\zeta_{0}\zeta_{1}}(\zeta ) & = & \frac{2}{2k-1}\frac{
(1-|\zeta_{0}|^{2})^{k}(1-|\zeta_{1}|^{2})^{k}} {(1-\zeta_{0}\zeta
_{1}^{\ast})^{2k}}    \label{Wksiksi} \nonumber \\
&\times& \int_{0}^{\infty}\rmd\lambda \lambda \tanh \left( \pi \lambda
\right) \Phi_{k}^{\frac{1}{2}}(\lambda )P_{-\frac{1}{2}+\rmi\lambda}\left( 
\frac{2\left( 1-\zeta^{\ast}\zeta_{0}\right) \left( 1-\zeta
_{1}^{\ast}\zeta \right)}{(1-|\zeta |^{2})(1-\zeta_{0}\zeta_{1}^{\ast})}
-1\right) \, .
\end{eqnarray}
\newpage


\begin{thebibliography}{99}
\bibitem{Hillery:1984aa} M.~Hillery, R.~F. O'Connell, M.~O. Scully, and
E.~P. Wigner. \newblock Distribution functions in physics: Fundamentals. %
\newblock {\em Phys. Rep.}, 106(3):121--167, 1984.

\bibitem{Lee:1995aa} H.-W. Lee. \newblock Theory and application of the
quantum phase-space distribution functions. \newblock {\em Phys. Rep.},
259(3):147--211, 1995.

\bibitem{Schroek:1996aa} F.~E. Schroek. 
\newblock {\em Quantum Mechanics on
Phase Space}. \newblock Kluwer, Dordrecht, 1996.

\bibitem{Ozorio:1998aa} A.~M. Ozorio~de Almeida. \newblock The {W}eyl
representation in classical and quantum mechanics. \newblock {\em Phys. Rep.}%
, 295(6):265--342, 1998.

\bibitem{Schleich:2001aa} W.~P. Schleich. 
\newblock {\em Quantum Optics in
Phase Space}. \newblock Wiley-VCH, Berlin, 2001.

\bibitem{QMPS:2005aa} C.~K. Zachos, D.~B. Fairlie, and T.~L. Curtright,
editors. \newblock {\em Quantum Mechanics in Phase Space}. \newblock World
Scientific, Singapore, 2005.

\bibitem{Polkovnikov:2010aa} A.~Polkovnikov. \newblock Phase space
representation of quantum dynamics. \newblock {\em Ann. Phys.},
325(8):1790--1852, 2010.

\bibitem{Weinbub:2018aa} J.~Weinbub and D.~K. Ferry. \newblock Recent
advances in Wigner function approaches. 
\newblock {\em Applied Physics
Reviews}, 5(4):041104, 2018.

\bibitem{Glauber:1963aa} R.~J. Glauber. \newblock Coherent and incoherent
states of the radiation field. \newblock {\em Phys. Rev.},
131(6):2766--2788, 1963.

\bibitem{Sudarshan:1963aa} E.~C.~G. Sudarshan. \newblock Equivalence of
semiclassical and quantum mechanical descriptions of statistical light
beams. \newblock {\em Phys. Rev. Lett.}, 10(7):277--279, 1963.

\bibitem{Agarwal:1968aa} G.~S. Agarwal and E.~Wolf. \newblock Quantum
dynamics in phase space. \newblock {\em Phys. Rev. Lett.}, 21(3):180--183,
1968.

\bibitem{Cahill:1969aa} K.~E. Cahill and R.~J. Glauber. \newblock Density
operators and quasiprobability distributions. \newblock {\em Phys. Rev.},
177(5):1882--1902, 1969.

\bibitem{Agarwal:1970aa} G.~S. Agarwal and E.~Wolf. \newblock Calculus for
functions of noncommuting operators and general phase-space methods in
quantum mechanics. I. mapping theorems and ordering of functions of
noncommuting operators. \newblock {\em Phys. Rev. D}, 2(10):2161--2186, 1970.

\bibitem{Gadella:1995aa} M.~Gadella. \newblock Moyal formulation of quantum
mechanics. \newblock {\em Fortschr. Phys.}, 43(3):229--264, 1995.

\bibitem{Agarwal:1981aa} G.~S. Agarwal. \newblock Relation between atomic
coherent-state representation, state multipoles, and generalized phase-space
distributions. \newblock {\em Phys. Rev. A}, 24(6):2889--2896, 1981.

\bibitem{Varilly:1989aa} J.~C. Varilly and J.~M. Gracia-Bond{\'{\i}}a. %
\newblock The Moyal representation for spin. \newblock {\em Ann. Phys.},
190(1):107--148, 1989.

\bibitem{Klimov:2010aa} A.~B. Klimov and H.~de~Guise. \newblock General
approach to {$\mathfrak{SU}(n)$} quasi-distribution functions. \newblock
\emph{J. Phys. A: Math. Theor.}, 43(40):402001, 2010.

\bibitem{Tilma:2016aa} T.~Tilma, M.~J. Everitt, J.~H. Samson, W.~J. Munro,
and K.~Nemoto. \newblock Wigner functions for arbitrary quantum systems. %
\newblock {\em Phys. Rev. Lett.}, 117(18):180401, 2016.

\bibitem{Gadella:1991aa} M.~Gadella, M.~A. Martin, L.~M. Nieto, and M.~A.
del Olmo. \newblock The {Stratonovich--Weyl} correspondence for one dimensional kinematical groups. \newblock {\em J. Math. Phys.},
32(5):1182--1192, 1991.

\bibitem{Nieto:1998cr} L.~M. Nieto, N.~M. Atakishiyev, S.~M. Chumakov, and
K.~B. Wolf. \newblock Wigner distribution function for {E}uclidean systems. %
\newblock {\em J. Phys. A: Math. Gen.}, 31(16):3875--3895, 1998.

\bibitem{Plebanski:2000fk} J.~F. Pleba{\'{n}}ski, M.~Prazanowski, J.~Tosiek,
and F.~K. Turrubiates. \newblock Remarks on deformation quantization on the
cylinder. \newblock {\em Acta Phys. Pol. B}, 31:561--587, 2000.

\bibitem{Kastrup:2006cr} H.~A. Kastrup. \newblock Quantization of the
canonically conjugate pair angle and orbital angular momentum. \newblock
\emph{Phys. Rev. A}, 73:052104, 2006.

\bibitem{Rigas:2011by} I.~Rigas, L.~L. S{\'a}nchez-Soto, A.~B. Klimov, J.~%
\v{R}eh\'a\v{c}ek, and Z.~Hradil. \newblock Orbital angular momentum in
phase space. \newblock {\em Ann. Phys.}, 326(2):426--439, 2011.

\bibitem{Kastrup:2016aa} H.~A. Kastrup. \newblock Wigner functions for the
pair angle and orbital angular momentum. \newblock {\em Phys. Rev. A},
94(6):062113, 2016.

\bibitem{Moyal:1949aa} J.~E. Moyal. \newblock Quantum mechanics as a
statistical theory. \newblock {\em Proc. Camb. Phil. Soc.}, 45(1):99--124,
1949.

\bibitem{Groenewold:1946aa} H.~J. Groenewold. \newblock On the principles of elementary quantum mechanics. \newblock {\em Physica}, 12(7):405--460, 1946.

\bibitem{Stratonovich:1956aa} R.~L. Stratonovich. \newblock On distributions
in representation space. \newblock {\em JETP}, 31:1012--1020, 1956.

\bibitem{Onofri:1975aa} E.~Onofri. \newblock A note on coherent state
representations of lie groups. \newblock {\em J. Math. Phys.},
16(5):1087--1089, 1975.

\bibitem{Perelomov:1986ly} A.~Perelomov. 
\newblock {\em Generalized Coherent
States and their Applications}. \newblock Springer, Berlin, 1986.

\bibitem{Zhang:1990aa} W.-M. Zhang, D.~H. Feng, and R.~Gilmore. \newblock %
Coherent states: Theory and some applications. 
\newblock {\em Rev. Mod.
Phys.}, 62(4):867--927, 1990.

\bibitem{Gazeau:2009aa} J.~P. Gazeau. 
\newblock {\em Coherent States in
Quantum Physics}. \newblock Wiley-VCH, Berlin, 2009.

\bibitem{Husimi:1940aa} K.~Husimi. \newblock Some formal properties of the
density matrix. \newblock {\em Proc. Phys. Math. Soc. Jpn.}, 22(4):264--314,
1940.

\bibitem{Kano:1965aa} Y.~Kano. \newblock A new phase-space
distribution function in the statistical theory of the electromagnetic
field. \newblock {\em J. Math. Phys.}, 6(12):1913--1915, 1965.

\bibitem{Berezin:1975mw} F.~A. Berezin. \newblock General concept of
quantization. \newblock {\em Commun. Math. Phys.}, 40:153--174, 1975.

\bibitem{Klimov:2017aa} A.~B. Klimov, J.~L. Romero, and H.~de~Guise. %
\newblock Generalized {SU(2)} covariant Wigner functions and some of their
applications. \newblock {\em J. Phys. A: Math. Theor.}, 50(32):323001, 2017.

\bibitem{Valtierra:2017aa} I.~F. Valtierra, J.~L. Romero, and A.~B. Klimov. %
\newblock {TWA} versus semiclassical unitary approximation for spin-like
systems. \newblock {\em Ann. Phys.}, 383:620--634, 2017.

\bibitem{Fano:1959ly} U.~Fano and G.~Racah. 
\newblock {\em Irreducible
Tensorial Sets}. \newblock Academic Press, New York, 1959).

\bibitem{Brif:1999kx} C.~Brif and A.~Mann. \newblock Phase-space formulation
of quantum mechanics and quantum-state reconstruction for physical systems
with lie-group symmetries. \newblock {\em Phys. Rev. A}, 59:971--987, 1999.

\bibitem{Lindblad:1970aa} G.~Lindblad and B.~Nagel. \newblock Continuous
bases for unitary irreducible representations of su(1,1). 
\newblock {\em
Ann. I. H. Poincare A}, 13:27--56, 1970.

\bibitem{Repka:1978aa} J.~Repka. \newblock Tensor products of unitary
representations of $sl_2(\mathbb{R})$. \newblock {\em Am. J. Math.},
100:747--774, 1978.

\bibitem{Holman:1966aa} W.~J. Holman and L.~C. Biedenharn. \newblock Complex
angular momenta and the groups su(1, 1) and su(2). \newblock {\em Ann. Phys.}%
, 39(1):1--42, 1966.

\bibitem{Wang:1970aa} K.-H. Wang. \newblock Clebsch-Gordan series and the Clebsch-Gordan coefficients of o(2, 1) and
su(1, 1). \newblock {\em J. Math. Phys.}, 11(7):2077--2095, 1970.

\bibitem{Orowski:1990aa} A.~Or{\l}owski and K.~W{\'o}dkiewicz. \newblock On
the {SU(1, 1)} phase-space description of reduced and squeezed quantum
fluctuations. \newblock {\em J. Mod. Opt.}, 37(3):295--301, 1990.

\bibitem{Brif:1997aa} C.~Brif. \newblock Su(2) andsu(1,1) algebra
eigenstates: A unified analytic approach to coherent and intelligent states. %
\newblock {\em Int. J. Theo. Phys.}, 36(7):1651--1682, 1997.

\bibitem{Kastrup:2003aa} H.~A. Kastrup. \newblock Quantization of the
optical phase space \newblock {\em Fortschr. Phys.}, 51(10-11):975--1134, 2003.

\bibitem{Seyfarth:2020aa} U.~Seyfarth, A.~B. Klimov, H.~de Guise, G.~Leuchs,
and L.~L. Sanchez-Soto. \newblock Wigner function for {SU}(1,1). \newblock
\emph{Quantum}, 4:317, 2020.

\bibitem{Figueroa:1990aa} H.~Figueroa, J.~M. Gracia-Bond{\'\i}a,
and J.~C. V{\'a}rilly. \newblock Moyal quantization with compact symmetry
groups and noncommutative harmonic analysis. \newblock {\em J. Math. Phys.},
31(11):2664--2671, 1990.

\bibitem{Wodkiewicz:1985aa} K.~Wodkiewicz and J.~H. Eberly. \newblock %
Coherent states, squeezed fluctuations, and the {SU(2)} and {SU(1,1)} groups
in quantum-optics applications. \newblock {\em J. Opt. Soc. Am. B},
2(3):458--466, 1985.

\bibitem{Gerry:1985aa} C.~C. Gerry. \newblock Dynamics of {SU(1,1)} coherent
states. \newblock {\em Phys. Rev. A}, 31(4):2721--2723, 1985.

\bibitem{Gerry:1991aa} C.~C. Gerry. \newblock Correlated two-mode {SU(1, 1)}
coherent states: nonclassical properties. \newblock {\em J. Opt. Soc. Am. B}%
, 8(3):685--690, 1991.

\bibitem{Gerry:1995kq} C.~C. Gerry and R.~Grobe. \newblock Two-mode
intelligent {SU(1,1)} states. \newblock {\em Phys. Rev. A},
51(5):4123--4131, 1995.

\bibitem{Jing:2011aa} J.~Jing, C.~Liu, Z.~Zhou, Z.~Y. Ou, and W.~Zhang. %
\newblock Realization of a nonlinear interferometer with parametric
amplifiers. \newblock {\em Appl. Phys. Lett.}, 99(1):011110, 2011.

\bibitem{Hudelist:2014aa} F.~Hudelist, J.~Kong, C.~Liu, J.~Jing, Z.~Y. Ou,
and W.~Zhang. \newblock Quantum metrology with parametric amplifier-based
photon correlation interferometers. \newblock {\em Nat. Commun.}, 5:3049,
2014.

\bibitem{Yurke:1986yg} B.~Yurke, S.~L. McCall, and J.~R. Klauder. \newblock
{SU(2)} and {SU(1,1)} interferometers. \newblock {\em Phys. Rev. A},
33(6):4033--4054, 1986.

\bibitem{Chekhova:2016aa} M.~V. Chekhova and Z.~Y. Ou. \newblock Nonlinear
interferometers in quantum optics. \newblock {\em Adv. Opt. Photon.},
8(1):104--155, 2016.

\bibitem{Li:2016aa} D.~Li, B.~T. Gard, Y.~Gao, C.-H. Yuan, W.~Zhang, H.~Lee,
and J.~P. Dowling. \newblock Phase sensitivity at the {H}eisenberg limit in
an {SU(1,1)} interferometer via parity detection. 
\newblock {\em Phys. Rev.
A}, 94(6):063840, 2016.

\bibitem{Banerji:1999aa} J.~Banerji and G.~S. Agarwal. \newblock Revival and
fractional revival in the quantum dynamics of su(1,1) coherent states. %
\newblock {\em Phys. Rev. A}, 59(6):4777--4783, 1999.

\bibitem{Erdelyi:1955aa} A.~Erd{\'e}lyi, W.~Magnus, F.~Oberhettinger, and
F.G Tricomi. \newblock {\em Higher Transcendental Functions}, volume~I. %
\newblock McGraw-Hill, New York, 1955.

\bibitem{NIST:DLMF} {NIST Digital Library of Mathematical Functions}. %
\newblock http://dlmf.nist.gov/, Chap.14, 2019. \newblock {F}.~W.~J. Olver,
A.~B. {Olde Daalhuis}, D.~W. Lozier, B.~I. Schneider, R.~F. Boisvert, C.~W.
Clark, B.~R. Miller, B.~V. Saunders, H.~S. Cohl, and M.~A. McClain, eds.

\bibitem{Alonso:2002aa} M.~A. Alonso, G.~S. Pogosyan, and K.~B. Wolf. %
\newblock Wigner functions for curved spaces. i. on hyperboloids. \newblock
\emph{J. Math. Phys.}, 43(12):5857--5871, 2002.
\end{thebibliography}


\end{document}